\documentclass[aps,superscriptaddress,preprint,showpacs]{revtex4}
\usepackage{mathrsfs}
\usepackage{amsmath}
\usepackage{graphicx}

\newcommand{\sech}{\mathrm{sech}}

\begin{document}

\title{Quantum electrodynamical shocks and solitons \\ in astrophysical plasmas}

\author{M.\ Marklund}
\affiliation{Department of Physics, Ume{\aa} University, SE--901 87
  Ume{\aa}, Sweden}

\author{D.\ D.\ Tskhakaya}
\affiliation{Institute of Physics, Georgian Academy of Sciences, 380077
Tbilisi, Georgia} 
\affiliation{Institute of Theoretical Physics, University
of Innsbruck, A-6020 Innsbruck, Austria }

\author{P.\ K.\ Shukla}
\affiliation{Department of Physics, Ume{\aa} University, SE--901 87
  Ume{\aa}, Sweden} 
\affiliation{Institut f\"ur Theoretische Physik IV, Fakult\"at f\"ur Physik und 
Astronomie, Ruhr-Universit\"at Bochum, D-44780 Bochum, Germany}

\date{\today}

\begin{abstract}
The nonlinear propagation of low-frequency circularly polarized waves in a
magnetized dusty plasma is analyzed. It is found that wave steepening and 
shock formation can take place due to the presence of nonlinear quantum
vacuum effects, thus giving rise to ultra-intense electromagnetic shocks.
Moreover, it is shown that solitary wave structures are admitted even
under moderate astrophysical conditions. 
The results may have applications to astrophysical plasmas, as well as next
generation laser interactions with laboratory plasmas containing dust clusters.
\end{abstract}

\maketitle

\section{Introduction}

The extreme environment offered in many astrophysical situations,
such as supernovae and binary mergers, 
in many cases give rise to new and exciting physics, e.g.\
neutrino--plasma interactions and gravitational waves. 
Furthermore, the next generation laser-plasma systems promises
to take laboratory astrophysics into new realms, as tens to hundreds
of petawatts, or even zetawatts, may be generated in femtosecond
pulses \cite{Mourou-etal}. Thus, as observational astronomy and 
cosmology grows even more accurate, and laboratory techniques 
continue to develop at the current rate, one will be able to probe
new physics, such as neutron star equations of state \cite{Shibata}
or the Unruh--Hawking effect \cite{Chen-Tajima}.

Recently, it has been noted that the inclusion of quantum electrodynamics
(QED) into plasma theory may give rise to novel wave modes
\cite{Marklund-etala,Marklund-etalb,Stenflo-etal}, in particular in 
dusty plasmas \cite{Marklund-etalc}. These wave modes could propagate
in a variety of environments, such as supernovae remnants, interstellar 
clouds, or laser-dusty plasma systems. 

In Ref.\ \cite{Heyl-Hernquist} it was found, using the method of
characteristics, that ultra-strong magnetic fields may induce shock front
formation in electromagnetic radiation, analogous to hydrodynamic shocks.
A possible application to such a scenario is magnetars, where field strengths
of the order $10^{15}\,\mathrm{G}$ can be found. 

In the present paper, we show that the nonlinear propagation
of electromagnetic waves in a a weakly magnetized dusty plasma 
may give rise to wave steepening and shock 
formation. This effect is solely due to the novel QED effect of photon--photon
scattering in a nonlinear quantum vacuum. The shock front formation may give rise
to ponderomotive acceleration of charged
particles in, e.g.\ interstellar clouds. 
The results are discussed for astrophysical energy scales.

\section{Pulse evolution}

In Ref.\ \cite{Marklund-etalc} the dispersion relation 
\begin{eqnarray}
  \frac{k^{2} c^{2} }{\omega^{2} } \approx \frac{4\alpha}{45\pi}\left[\left( 
  \frac{E_0}{E_S} \right)^2 \frac{k^{2} c^{2} }{\omega^{2} } + \left(\frac{cB_0%
  }{E_S}\right)^2 \right]\frac{k^2c^2}{\omega^2} - \frac{\omega_{pd}^2\gamma_d%
  }{\omega_{cd}(\omega\gamma_d - \omega_{cd})} .  \label{eq:transverse}
\end{eqnarray}
was found. The relation (\ref{eq:transverse}) is valid for low-frequency
circularly electromagnetic polarized waves, with electric field $\mathbf{E} = E (\hat{%
\mathbf{x}} + i\hat{\mathbf{y}})\exp(ikz - i\omega t)$, propagating along a
magnetic field $\mathbf{B}_0 = B_0\hat{\mathbf{z}}$ Moreover, the critical
Schwinger field is given by $E_S = m_{0e}^2c^3/e\hbar \sim 10^{18}$ V/m
(see, e.g.\ Ref.\ \cite{Schwinger}), where $c$ is the speed of light in
vacuum, $\hbar$ is Planck's constant divided by $2\pi$, $m_{0e}$ is the
electron rest mass, $e$ is the absolute value of the electron charge, and $%
\omega_{cd} = q_dB_0/m_{0d}$ and $\omega_{pd} = (n_{0d} q_d^2/\epsilon_0
m_{0d})^{1/2}$ is the dust gyro and plasma frequency, respectively. Here $n_{0d}$ 
denotes the dust particle density in the laboratory frame, $%
q_{d} $ the dust charge, $m_{0d}$ the dust particle rest mass, and $\alpha $ is the
fine-structure constant.

Next, we introduce the dimensionless variables and parameters $\Omega =\omega /\omega
_{cd} $, $K=kc/\omega _{pd}$, and $\kappa = 
(4\alpha /45\pi)^{1/2}\omega _{pd}/\omega _{cd}$, and normalize the 
electric field according to $\mathcal{E} = E/E_S$.
We consider the case of
ultra-relativistic dust in a weak magnetic field, for which \cite{Marklund-etalc} 
\begin{equation}
  K^{2}\Omega ^{2}+\Omega ^{3}=\kappa ^{2}|\mathcal{E}|^{2}K^{4},  \label{eq:new}
\end{equation}
holds. 

The generic evolution equation for a slowly varying
envelope $\mathcal{E}$ following from the dispersion relation
(\ref{eq:new}) reads
\begin{equation}\label{eq:general}
  i\partial_t^3\mathcal{E} - \partial_t^2\partial_z^2\mathcal{E} 
  + \kappa^2|\mathcal{E}|^2\partial_z^4\mathcal{E} = 0 ,
\end{equation}
where $iK\rightarrow \partial _{z}$ and $i\Omega \rightarrow
-\partial _{t}$ has been used.

If the quantum parameter $\kappa \rightarrow 0$, from Eq.\ (\ref{eq:new}
we obtain the equation 
\begin{equation}
  (i\partial _{t}-\partial _{z}^{2})\mathcal{E}(t,z)=0,  \label{eq:prop}
\end{equation}
yielding a linear traveling wave solution with the whistler-like dispersion
relation $\Omega =-K^{2}$

In the case of small frequencies, $\Omega \ll K^{2},$ (the case with a
dominant nonlinear term), we neglect the third order frequency $\Omega ^{3}$
in Eq.\ (2) and obtain 
\begin{equation}
  \partial _{t}\mathcal{E}\pm \kappa |\mathcal{E}|\partial _{z}\mathcal{E} 
  = 0.  \label{eq:prop2}
\end{equation}
Thus, letting $\mathcal{E}(t,z)=|\mathcal{E}(t,z)|\exp [i\varphi (t,z)]$ 
the amplitude $|\mathcal{E}(t,z)|$ satisfies the equation 
\begin{equation}
  \partial _{t}|\mathcal{E}|\pm \kappa |\mathcal{E}|\partial _{z}|\mathcal{E}| 
  = 0,  \label{eq:amp}
\end{equation}
while the phase is connected with the amplitude by the equation 
$
\partial _{t}\varphi \pm \kappa |\mathcal{E}|\partial _{z}\varphi =0.
$ 
For simple waves \cite{Landau-Lifshitz} Eq.\ (\ref{eq:amp}) yields the solution 
\begin{subequations}
\begin{equation}
  z=\pm \kappa |\mathcal{E}|+f(|\mathcal{E}|),  \label{eq:steep1}
\end{equation}
i.e.\ 
\begin{equation}
  |\mathcal{E}(t,z)|=f^{-1}(z\mp \kappa |\mathcal{E}|t),  \label{eq:steep2}
\end{equation}
\end{subequations}
where the function $f^{-1}$ is the inverse function of $f$ and $f$ is
determined by the initial condition at $t=0$. Thus, we have to define the
initial condition $|\mathcal{E}(0,z)|=f(z)$ at $t=0$ and then 
find $z=f^{-1}(|\mathcal{E}|)$. As
Eq.\ (\ref{eq:steep2}) shows, the wave steepening due to nonlinearity  may
occur. This can also be seen via Eq.\ (\ref{eq:steep1}), from which we find that
\begin{equation}\label{eq:div}
  {\partial_z}|\mathcal{E}|=\frac{1}{\pm\kappa t+f^{^{\prime }}(|\mathcal{E}|)}.
\end{equation}
where $f^{\prime }(x) \equiv \partial_x f(x)$. 
\begin{enumerate}
  \item[(a)] If  $f^{\prime }(|\mathcal{E}|) < 0$ (which can be interpreted as the 
  initial function $|\mathcal{E}|=f(z)$ at $t=0$ decreasing with $z$), we have 
  at time $t=-|f^{\prime}(|\mathcal{E}|)|/\kappa $ that $\partial _{z}|\mathcal{E}| 
  = \infty$, if the plus is chosen in front of the $\kappa$-term in Eq.\ (\ref{eq:div}). 
  \item[(b)] If we, on the other hand, have $f'(|\mathcal{E}|) > 0$, we choose the
  minus sign in front of the $\kappa$-term in Eq.\ (\ref{eq:div}), and
  obtain the shock front formation for a finite time.
\end{enumerate}
Thus, we clearly have the
shock wave formation at this instant \cite{Landau-Lifshitz}. Starting
with a Gaussian pulse, case (a) implies the wave steepening and the shock
front formation in the leading part of the pulse, while case (b) implies
the wave steepening in the tail of the pulse, which can be seen by a simple
spatial reflection symmetry. We note that in principle 
density gradient effects in the dusty plasma should be taken into
account once significant steepening occurs, but this turns out
to be a higher order effect.

In the weakly nonlinear case, we may use Eq.\ (\ref{eq:prop})
to simplify the general pulse evolution equation (\ref{eq:general}),
using $\partial^2_z\mathcal{E} \approx i\partial_t\mathcal{E}$. 
Assuming a slowly varying envelope $\mathcal{E}$, Eq.\
(\ref{eq:general}) yields
\begin{equation}\label{eq:nlse}
  i\partial_t\mathcal{E} - \partial_z^2\mathcal{E} 
  - \kappa^2|\mathcal{E}|^2\mathcal{E} = 0 ,
\end{equation}
which is the well-known nonlinear Schr\"odinger equation
with a focusing nonlinearity. Thus, such an equation
admits the formation of a bright soliton \cite{Hasegawa} 
given by 
$  \mathcal{E}(t,z) = \mathcal{E}_0\,\sech(\mathcal{E}_0 z)%
  \exp\left(- {i\mathcal{E}_0^2 t}/{2\sqrt{2}} \right)$, 
where $\mathscr{E}_0$ is the amplitude, and we have assumed 
that the soliton has the peak value at $t = 0$ located at $z = 0$. 
Thus, in the weakly nonlinear case, a balance between the
dispersive and nonlinear effects may be struck. However, in practice,
the coherence length of such structures could be short, due to e.g.\ 
plasma inhomogeneities. Moreover, while the one-dimensional bright soliton 
is stable, higher dimensional effects may yield instabilities. 
It is interesting thought that such structures 
could form with rather modest parameter values, as will be discussed
next.

\section{Discussion and conclusions}

The effect of the wave steepening and intense shock front
formation presented here takes place whenever the 
electromagnetic intensity is high in the dusty plasma. Thus,
a natural question arises: What are the situations when this
may occur? As astrophysical systems offer extreme energy 
scales, these are natural places to look for QED effects, such as the
one presented here. Moreover, the presence of dusty plasmas
is well known in a wide range of astrophysical environments
\cite{Horanyi-etal,Okamoto-etal,Shukla-Mamun}, 
such as supernovae \cite{Dunne-etal,Krause-etal} and 
interstellar clouds \cite{Mendis-Rosenberg}. In the case of 
Interstellar clouds, we have an expected dust density of the order 
$n_{0d} \sim 10^{-1}\,\mathrm{m^{-3}}$ 
\cite{Mendis-Rosenberg}, with a typical dust mass $m_{0d} \sim 
10^6\,m_{\text{proton}} \sim 10^{-21}\,\mathrm{kg}$ \cite{Shukla-Mamun}. 
A value $B_0 \sim 10^{-8}\,\mathrm{G}$ (well within the limits for 
such environments) results in $\kappa \sim 10^5$. If the circularly polarized 
wave satisfy
$E_0/E_S \approx 2\times10^{-5}$, a value appropriate at a distance of 
1\,AU from a supernova, we have $\kappa|\mathcal{E}| \sim 1$, thus
giving an appreciable nonlinearity in Eqs.\ (\ref{eq:amp}) and (\ref{eq:nlse}).

In Fig.\ 1 we have plotted the pulse evolution in two different cases.
The upper panel represents the positive sign of Eq.\ (\ref{eq:amp}),
while the lower panel depicts the pulse evolution for the negative 
sign in Eq.\ (\ref{eq:amp}). We have used the initial pulse amplitude
$|\mathcal{E}(0, z)| = \mathcal{E}_0 \exp[-(z - z_0)^2/2a^2]$, with $z_0 = 3$ and $a = 1$.
Moreover, the initial amplitude satisfies $\kappa \mathcal{E}_0 = 0.7$, thus
well within the limits posed by, e.g.\ interstellar clouds at 1\,AU from
a supernova. We have plotted the evolution of $|\mathcal{E}(t, z)|/\mathcal{E}_0$. The effect
of the nonlinearity can clearly be seen in the form of the wave steepening.

One interesting aspect of the shock front formation is the possibility
of charged particle acceleration due to the ponderomotive force. However, 
before conclusions can be drawn on possible particle acceleration, analysis 
of mechanisms rendering the pulse wave front less steep must be made.

\acknowledgments

This research was partially supported by the Swedish Research Council
through the contract No. 621-2004-3217.

\newpage

\newpage

\begin{figure}
  \includegraphics[width=.8\textwidth]{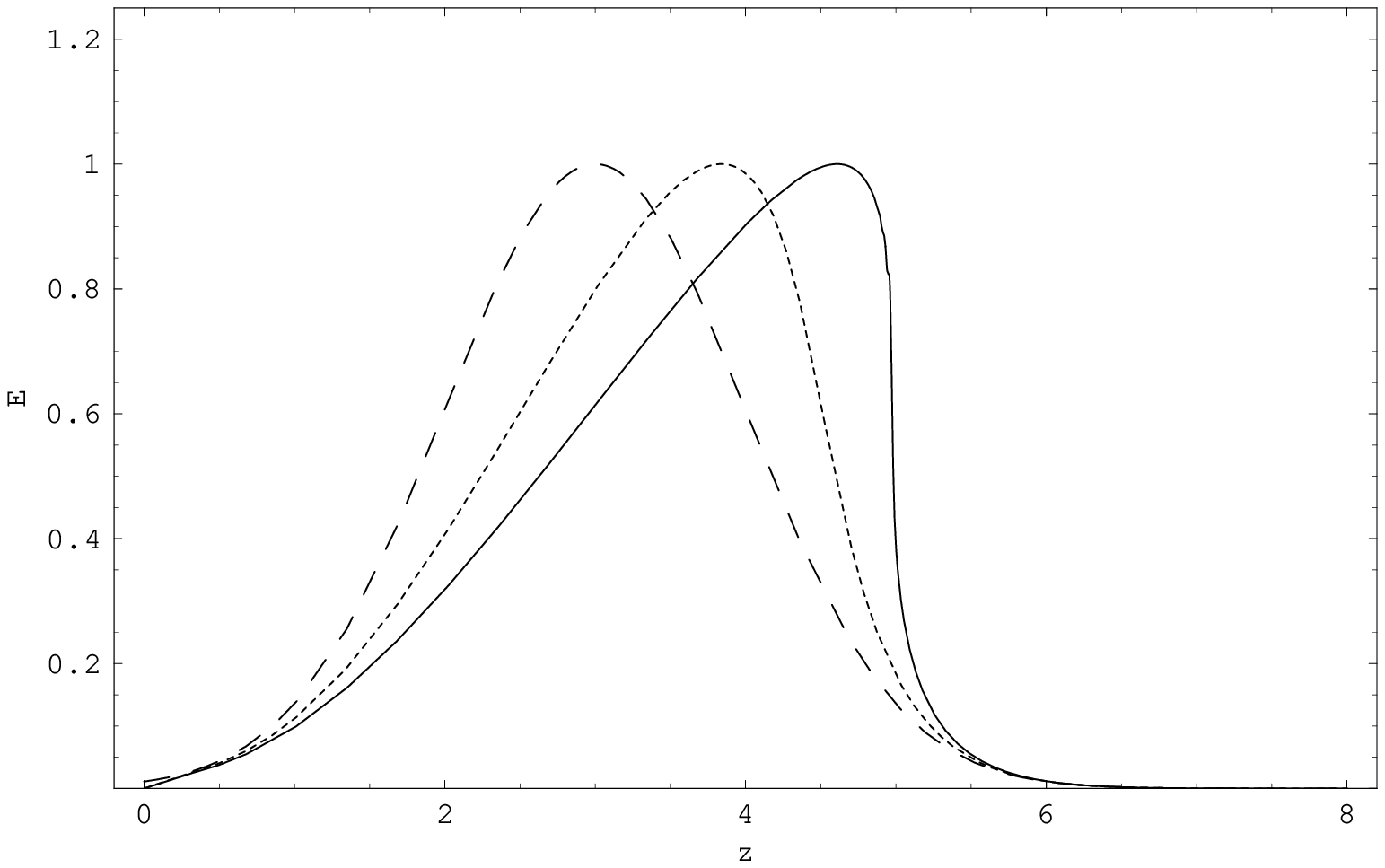}
  \includegraphics[width=.8\textwidth]{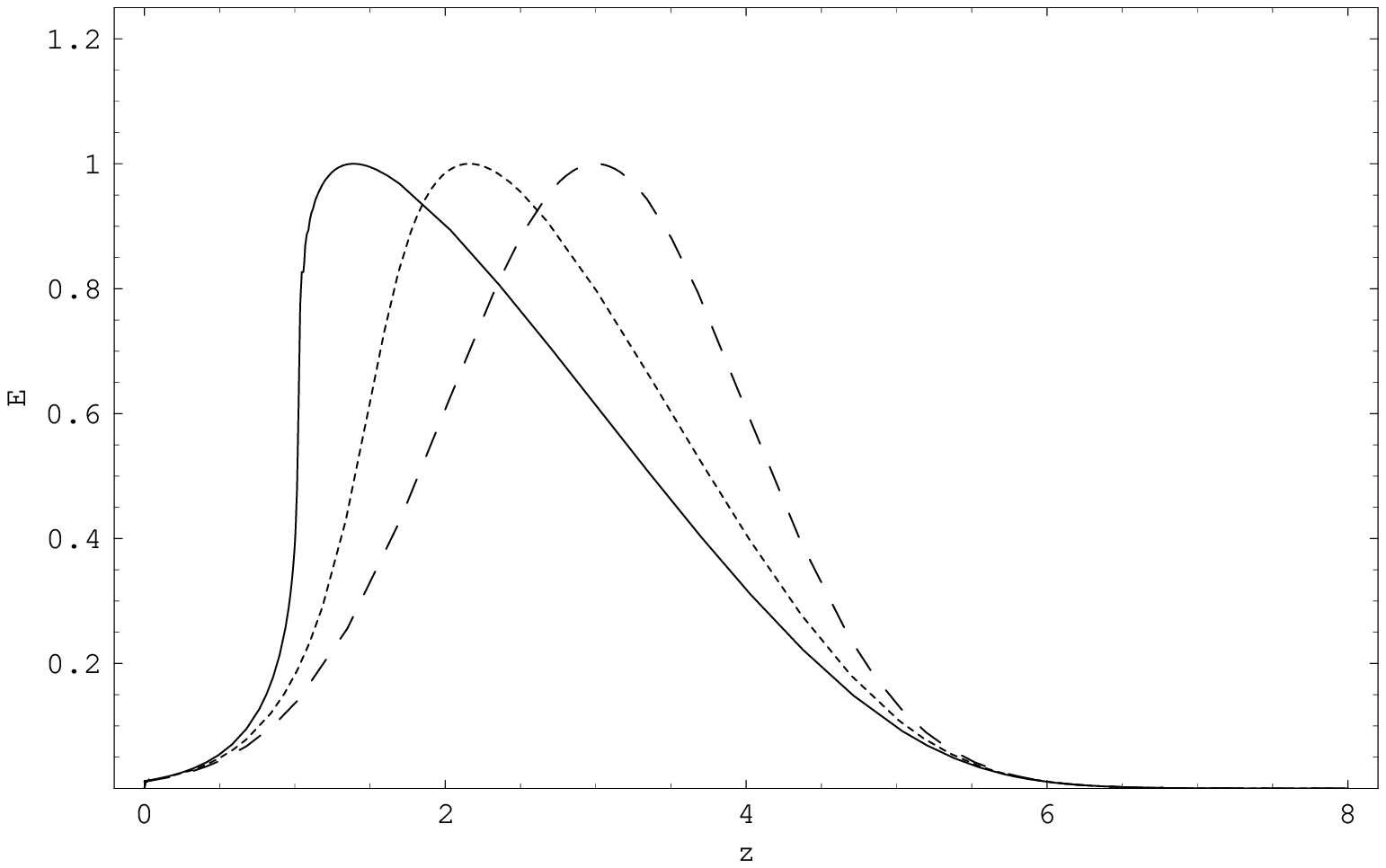}
  \caption{The magnitude $|\mathcal{E}|$ of the normalized electric 
  field relative to the initial value
    $\mathcal{E}_0$ plotted as a function of $z$ for three different times. 
    At $t = 0$ we have the dashed 
    Gaussian curve, while at $t = 12$ and $t = 23$ we have the dotted
    and full curve, respectively. In the upper panel we have used the 
    positive sign in Eq.\ (\ref{eq:amp}). The wave 
    steepening and shock front formation can clearly be seen. In
    the lower panel, we have instead used the negative sign in Eq.\
    (\ref{eq:amp}). As expected, the wave 
    steepening and shock front formation now occurs at
    the trailing edge of the pulse.}
\end{figure}

\end{document}